\documentclass[11pt]{article}
\usepackage{geometry}    % See geometry.pdf to learn the layout options. There are lots.
\usepackage{fancyhdr} % Not a default package
\usepackage{latexsym}
\usepackage{graphics}
\usepackage{graphicx}
\usepackage{epstopdf}
\usepackage{epsfig}
\usepackage{amssymb}
\usepackage{makeidx}
\usepackage{amsfonts}
\usepackage{amstext}
\usepackage{amsmath}
\usepackage{amsbsy}
\usepackage{wasysym}
\usepackage{mathrsfs}
\usepackage{authblk}
\newlength{\InitialBLS}
\author[*]{E. Segreto}
\affil[ ]{INFN - Laboratori Nazionali del Gran Sasso, Assergi, Italy}

\date{}                                           % Activate to display a given date or no date

\title{An analytic technique for the estimation of the light yield of a scintillation detector.}

\begin{document}
\maketitle
\let\oldthefootnote\thefootnote
\renewcommand{\thefootnote}{\fnsymbol{footnote}}
\footnotetext[1]{E-mail: ettore.segreto@lngs.infn.it}
\let\thefootnote\oldthefootnote
\begin{abstract}
A simple model for the estimation of the light yield of a scintillation detector is developed under general assumptions and relying exclusively on the knowledge of its optical 
properties. The model allows to easily incorporate effects related to Rayleigh scattering and absorption of the photons.
The predictions of the model are benchmarked with the outcomes
of Monte Carlo simulations of specific scintillation detectors. An accuracy at the level of few percent is achieved.
The case of a real liquid argon based detector  is explicitly treated and the
predicted light yield is compared with the measured value.
\end{abstract}

\section{Introduction.}
 A typical scintillation detector is constituted by a scintillating material contained in a reflective box and  by a system of one or more Photo-Sensitive Devices (PSDs) that 
\emph{observes} the active medium. This 
kind of detectors is widely used in many fields of physics to detect particles or energetic photons (\cite{leo}, \cite{knoll}, \cite{birks}).  
%In many cases they are capable to measure the
%energy that the ionizing radiation leaves inside the active medium.  
In many cases they also allow to perform calorimetric measurements since  their output signal is often proportional to the energy that the ionizing radiation leaves inside the 
active medium.
The constant ratio between the signal (usually in charge) and the deposited energy is the light yield (LY)  and is  measured in (photo-) electrons/keV.  The main factors that 
determine the LY of a detector  are the abundance of photons produced per unit of deposited energy  (photon yield), the optical properties of the scintillator and of the internal 
surface of the box where it is contained, the number and dimensions of the PSDs and their efficiency in converting photons into a detectable signal. The LY is one of the parameters 
that more deeply influences the design of a scintillation detector since, for each value of deposited energy, it fixes the scale of the intensity of the output signal. Therefore the layout of the detector needs to be optimized in order to match as well as possible the  LY with the
 characteristics (energy, Linear Energy Transfer, ...) of the incoming radiation and of the electronic read-out chain. In the field of 
low energy particle physics (neutrino physics, direct Dark Matter search, double $\beta$ decay experiments, ...), for instance, where high sensitivity and high energy resolution is 
required the optimal LY value is often the highest achievable, compatibly with the others experimental constraints.
% in order to lower as much as possible the detection 
%threshold for rare events and to have the highest energy resolution the optimal value of the LY is often the highest achievable, compatibly with the constraints that physics imposes 
%to the detector and with the available resources. 
Traditionally the LY is evaluated by means of Monte Carlo simulations. This technique has the advantage of giving 
very precise results  at the expense of programming and mainly running  codes that typically invoke the propagation of millions of photons and  this 
can result extremely cumbersome especially in an optimization process. In this work an alternative and completely analytic approach for the estimation of the LY of simple detectors 
is presented. It offers the possibility to obtain fast and robust results, with an accuracy of few per cent if compared with the classical Monte Carlo approach. Furthermore the explicit 
dependence of the LY from all the optical parameters makes this technique particularly suitable for the process of detector optimization.

% In this work the LY is estimated through an analytic technique that simply requires the knowledge of the aforementioned  parameters.

%In this work it is illustrated an analytic way to estimate the LY of a scintillation detector starting from the knowledge of the optical parameters of the
%materials it is made of and of the characteristics of the PSDs.\\
%The LY is one of the most important parameters that drives the design of a scintillation detector since it determines its performances in a certain range of deposited energy. For this
%reason having a tool that allows to optimize it with respect to the optical properties of the passive and active components of the detector (reflectivity, efficiency and dimensions of the
% PSDs, ...) can be very helpful in many practical applications.
 
\section{Light Yield calculation.}
\label{sec:basic}
This work addresses the most simple and common layout for a scintillation detector: an uniform scintillating medium (in liquid, gaseous or solid state) is contained in a cell with 
(highly) reflecting  internal surfaces and is observed by one ore more PSDs whose window(s) are installed on the internal surface of the cell. The following hypotheses are 
assumed to hold:
\begin{itemize}
\item The scintillator is uniform and completely fills the cell where it is contained. It is monochromatic. This is not a limitation for the model because in the case of a non
 monochromatic scintillator one can calculate the LY as a function of the wavelength, $\lambda$, of the emitted radiation and average it over the entire spectrum, that is:
 \begin{equation}
 LY=\frac{\int_0^{\infty}LY(\lambda)I(\lambda)~d\lambda}{\int_0^{\infty}I(\lambda)~d\lambda}
 \end{equation}
 The scintillator is initially considered to be  perfectly transparent to scintillation radiation, that is photons do not suffer absorption or elastic scattering (Rayleigh scattering) processes 
 along their propagation. This assumption will be removed in section \ref{sec:ray_abs}.
 \item The LY does not depend on the point where the ionizing radiation leaves its energy. This implies that the scintillating medium is contained in a cell of regular shape (sphere, 
 square cylinder, cube...) and that its internal surface is reflective and its reflectivity R is high.
 \item Each PSD is schematized as: a window with given optical optical properties (reflectivity and transmissivity) coupled to a \emph{"black box"} that absorbs all transmitted 
 photons 
 and produces a signal with a certain efficiency.\footnote{In the case of a photomultiplier this is not directly the Quantum Efficiency, because the latter includes in its definition effects 
 related to window reflectivity and transmissivity and photo-cathode reflectivity. In the present schematization these effects are all transferred to the window and the PSD 
 (photomultiplier) efficiency is the probability that a photon absorbed by the photo-cathode (black box) produces a photo-electron.}  
\end{itemize}
The reflectivity of the internal surface of the cell and of the PSD window  (and consequently its transmissivity) typically depends on the angle of impact, $\theta$, of the photon with
 respect to the normal to the surface. In this work only average values are considered. Assuming an uniform lambertian illumination the average reflectivity of a flat infinitesimal 
 element of surface can be calculated as:
 \begin{equation}
 R = \int_{0}^{\pi /2} R(\theta)~2~ sin\theta cos\theta ~d\theta
 \end{equation}
 the same holds for the transmissivity.\\
 
 \subsection{Basic calculation.}
The LY of this kind of scintillation detector can be factorized into three terms:
\begin{equation}
LY = N_{\gamma} \times \epsilon_{opt}  \times \epsilon_{PSD}
\label{eq:ly_bare}
\end{equation}
where:
\begin{itemize}
\item  $N_{\gamma}$ is the \emph{photon yield} of the scintillator: it is the number of photons produced per
unit of deposited energy by a certain radiation (usually in photons/keV);
\item  $\epsilon_{opt}$ is the \emph{optical efficiency}: it is the fraction of the originally produced photons that  manages to cross the windows of the PSDs. It depends on the optical 
properties of the boundary surface of the detector, of the scintillation medium and of the PSDs' windows;
\item $\epsilon_{PSD}$ is the \emph{conversion efficiency} of the PSDs: it is the efficiency of the PSD system in converting photons into signal (\emph{photo-electrons}). 
\end{itemize}
The dimensions of the $LY$ are \emph{photo-electrons/keV}. $N_{\gamma}$ and  $\epsilon_{PSD}$ are characteristic parameters of the scintillator medium and of the photo-sensitive 
devices and in the majority of the cases are precisely known. On the other side $\epsilon_{opt}$ is typically unknown and needs to be estimated. It represents the average 
probability that a scintillation photon  produced in the active medium by an energy release reaches and crosses the window of one of the PSDs, surviving to the processes that can kill it
while bunching inside the detector.\\

The propagation of photons inside a scintillation detector is an intrinsically recursive process. Consider, for example, a sphere containing a scintillating medium and assume that a 
fraction $f$ of its internal surface is occupied by the window of a PSD, that is perfectly transparent to scintillation radiation and with refractive index matched to that of the scintillator. 
Neglect, for now, absorption and elastic scattering phenomena.
%We neglect Rayleigh scattering and absorption. 
A photon produced in a random point inside the sphere and with a random direction when reaches the boundary surface has an 
average probability $f$ to be detected  (assuming  $\epsilon_{PSD}=1$), since its impact point is uniformly distributed on the sphere. On the other hand it has a probability $(1-f)$ to
 hit the non active surface and if its reflectivity $R$ is not zero it is sent back inside the scintillator with probability $R(1-f)$. Reflected photon has again a random direction and  a
random production point (on the surface of the sphere this time) and has again a detection probability equal to $f$ and a probability to be reflected equal to $R(1-f)$. The same
 situation will repeat again identical to itself after any reflection.\\
 
Let's generalize these ideas and consider a general scintillation detector. In order to estimate its optical efficiency $\epsilon_{opt}$ assume that the process that starts with the
production and ends with the absorption/detection of the photon can be treated in a recursive way. This means that it can be divided  into a series of subsequent and 
indistinguishable steps and that it is possible to define two quantities, $\alpha$ and $\beta$, where:
\begin{itemize}
\item{$\alpha$ is the \emph{average}  probability \emph{per step} that a photon randomly generated in the scintillator 
volume (for the first step) or surviving from the previous step is {\textbf{detected}}.\footnote{Hereafter for \emph{detected} photons we mean photons that succeed in 
crossing the PSD window.} }
\item{$\beta$ is the \emph{average}  probability \emph{per step} that a photon is {\textbf {regenerated}}, that is the probability that it is not \emph{lost} (detected or  absorbed)
and that some  physical process randomizes again its direction (reflection for instance).}
\item{$\alpha$ and $\beta$ are constant for all the steps (from the assumption on the recursiveness of the process).}
\item{$\alpha\le 1$ and $\beta<1$.}
\end{itemize}
With these assumptions it is easy to calculate the detection and regeneration probabilities for a photon at step $n$ \emph{after  surviving to the previous $n-1$ steps}. The values
 are
shown in table \ref{tab:detreg}.
%***************************** TABLE 1 *****************************% 
\begin{table}[htbp]
\caption{\textsf{\textit{Detection and regeneration probabilities for a photon propagating inside a scintillation detector as a function of the propagation step (see text).}}}
\begin{center}
\vspace{0.5cm}
\begin{tabular}{l|c|c} 
& detection probability & regeneration probability \\
\hline  
step 0 & $\alpha$ & $\beta$ \\
\hline
step 1 & $\alpha\beta$ & $\beta^2$\\
\hline
step 2 &  $\alpha\beta^2$ & $\beta^3$\\
\hline
$\dots$ & $\dots$ & $\dots$ \\
\hline
step n & $\alpha\beta^n$ & $\beta^n$ \\
\hline
\hline
\end{tabular}
\label{tab:detreg}
\end{center}
\end{table}
%***********************************************************************%
Hence the optical efficiency, that is the sum of these detection probabilities over all the steps can be simply calculated as the sum of a geometric series: 
\begin{equation}
\label{eq:mas}	
\epsilon_{opt} = F(1,\alpha,\beta) = \sum_{n=0}^{\infty} \alpha{\beta}^n = \frac{\alpha}{1-\beta}
\end{equation}
This series converges because $\beta < 1$. The notation $F(Q,\alpha,\beta)$ will be fully clear in the next section. In principle, for each given detector, one could define the
 elementary step in many different ways and this can make the calculations more or less difficult, but the final result is general and absolutely independent of the step definition.\\

Consider again the simple spherical scintillation detector described above. The step can be defined in a natural way as  the photon propagation between subsequent interactions
with the boundary surface: it starts just after one reflection and ends when photon hits again the detector's walls. With this step definition $\alpha$ e $\beta$ are easily calculable, in
fact the photon will have at each step:
\begin{itemize}
\item a probability $f$ to be detected $\Rightarrow \alpha=f$;
\item a probability $1-f$ to hit the non active internal surface and a probability  $R(1-f)$ to be regenerated $\Rightarrow \beta=R(1-f)$;
\end{itemize}
It is now possible to calculate the optical efficiency of the detector using equation \ref{eq:mas}:	
\begin{equation}
\label{eq:simple_scint}
\epsilon_{opt}=F(Q=1,f,R(1-f))=\frac{f}{1-R(1-f)}
\vspace{0.7cm}
\end{equation}
 
In a realistic situation the PSD window has non trivial optical characteristics, that is a transmissivity $T_{w}\ne 1$ and a reflectivity $R_{w}\ne 0.$
%\footnote{$T_w$ and $R_w$ can eventually include effects related to the non perfect matching on the refractive indexes between the scintillator and the PSD's window}
In this case one has:
 
\begin{eqnarray}
\label{eq:simple_1}
\alpha = T_{w}f~~~~~~~~~~~~~~~\nonumber\\
\nonumber\\
\beta =R(1-f)+R_{w}f\\
\nonumber\\
Q = 1~~~~~~~~~~~~~~~~~~~~\nonumber
\end{eqnarray}
 
where the term $R_{w}f$ in the definition of $\beta$ takes into account the regeneration probability on the PSD's window and:
 
\begin{equation}
\label{eq:simple_scint_1}
\epsilon_{opt}=F(Q=1,T_{w}f,R(1-f)+R_{w}f)=\frac{T_{w}f}{1-R(1-f)-R_{w}f}
\vspace{0.7cm}
\end{equation}

Equation \ref{eq:simple_scint_1} has been derived for a spherical detector, but it can be safely considered a very good approximation for all regular box shapes and, more 
generally, for all the  cases where $\alpha/T_w$ can be (roughly) identified with $f$ (PSD surface coverage).  

In a even more general experimental situation the scintillation detector could  host  more than one PSD. In this case one should consider one PSD per time and calculate the
optical efficiency (equation \ref{eq:simple_scint_1}) with respect to it. For the reflectivity of the remaining part  of the cell one should take the average reflectivity of the non 
active surface and of the remaining PSDs' windows, weighted by their relative surface coverage. The same procedure should be repeated for each one of the PSDs in the cell  and
the total optical efficiency is obtained by summing up all the individual efficiencies.\\

\subsection{Rayleigh scattering and photon absorption.}
\label{sec:ray_abs}
An ideal scintillation medium is perfectly transparent to its own radiation and thus the emitted photons propagate unabsorbed along straight  trajectories between a reflection and 
the other. In 
practice this is never the case and photons have a finite probability of being absorbed or elastically scattered (Rayleigh scattering) while traveling across the scintillator because of 
the presence of some contaminant(s) or because of the intrinsic molecular/atomic structure of the medium. The model developed above allows to  include these effects and 
disentangle them from reflections in an extremely clean way.  
%We now want to take explicitly into account the effects of Rayleigh scattering and of photon absorption and disentangle them from the effects of reflections on the boundary 
%surface of the cell. 
Consider a cell filled with an uniform scintillator. Assume that $\alpha_0$ and $\beta_0$ are the detection and regeneration probabilities calculated with respect to one possible 
step definition \emph{in absence of scattering/absorption}. If, instead, these effects are present, the step definition needs to be opportunely modified and the detection and 
regeneration
probabilities, $\alpha$ and $\beta$, recalculated. If the photon interacts in the scintillator it can be either scattered or absorbed. Absorption kills the photon while 
scattering \emph{regenerates} it more or less in the same way as a reflection does. Following this consideration the step definition needs only to be slightly enlarged to include the 
case 
that an interaction can stop the current step and start a new one. In order to calculate $\alpha$ e $\beta$ the following quantities need to be defined:
\begin{itemize}
%\item{the detection and the regeneration probabilities {\it{in absence of scattering and absorption}}: $\alpha_{0}$ and $\beta_{0}$;}
\item{the \emph{effective interaction} length, $\tilde{\lambda}$:
\begin{equation}
\frac{1}{\tilde{\lambda}} = \frac{1}{\lambda_{R}} +\frac{1}{\lambda_{A}}
\end{equation}
where $\lambda_{R}$ and $\lambda_{A}$ are the Rayleigh scattering length and the absorption length respectively; }
\item{the average probability, $U_{RA}$, that a photon randomly  produced inside the detector reaches the end of the step, as defined in \emph{absence} of scattering/absorption,
 without interactions}.
\end{itemize}
  
The detection and regeneration probabilities (per step)  of the photon are then:
\begin{eqnarray}
\alpha = U_{RA}\alpha_{0}~~~~~~~~~~~~~~~~~~~~~\nonumber\\
\nonumber\\
\beta= U_{RA}\beta_{0}+(1-U_{RA})\frac{\tilde{\lambda}}{\lambda_{R}}
\end{eqnarray}

In the definition of $\beta$ the term $U_{RA}\beta_{0}$ accounts for the probability that the photon reaches \emph{untouched} the boundary surface of the cell and is regenerated 
by 
reflection while the term  $(1-U_{RA})\frac{\tilde{\lambda}}{\lambda_{R}}$ accounts for the  probability that the photon interacts before reaching the boundary surface  and is  
regenerated by scattering.
  
From equation \ref{eq:mas} the optical effciency will be:

\begin{equation}
\epsilon_{opt}^{RA} = \frac{U_{RA}\alpha_{0}}{1-[U_{RA}\beta_0+(1-U_{RA})\frac{\tilde{\lambda}}{\lambda_{R}}]}
\end{equation}

after some algebra one finds that:

\begin{equation}
\label{eq:eopt_RA}
\epsilon_{opt}^{RA}=\frac{\alpha_{0}}{Q-\beta_{0}} 
\end{equation}

where:

\begin{equation}
\label{eq:Q}
Q=\frac{1-(1-U_{RA})\frac{\tilde{\lambda}}{\lambda_{R}}}{U_{RA}}
\end{equation}
Equation \ref{eq:eopt_RA} generalizes in an elegant way the case of photons traveling in a perfectly transparent  medium: all the effects related to scattering/absorption are 
contained in the term $Q$ and can be treated/calculated separately. For this reason it seems to be appropriate the notation $F(Q,\alpha_0,\beta_0)$, where all the three terms that 
contribute to the \emph{optical efficiency} are expressly indicated.\\  
The term $Q$ can be  calculated for the interesting case of the detector  described in section \ref{sec:basic}. One can write in an almost general way that:
\begin{equation}
U_{RA}=\int_{0}^{\infty} P(x)e^{-\frac{x}{\tilde{\lambda}}}dx\simeq \int_{0}^{\tilde{L}} P(x)e^{-\frac{x}{\tilde{\lambda}}}dx
\end{equation}

%*****************************FIGURE 1*****************************%
\begin{figure*}[tbp]
\begin{center}
\includegraphics*[width=10.cm,height=8.cm]{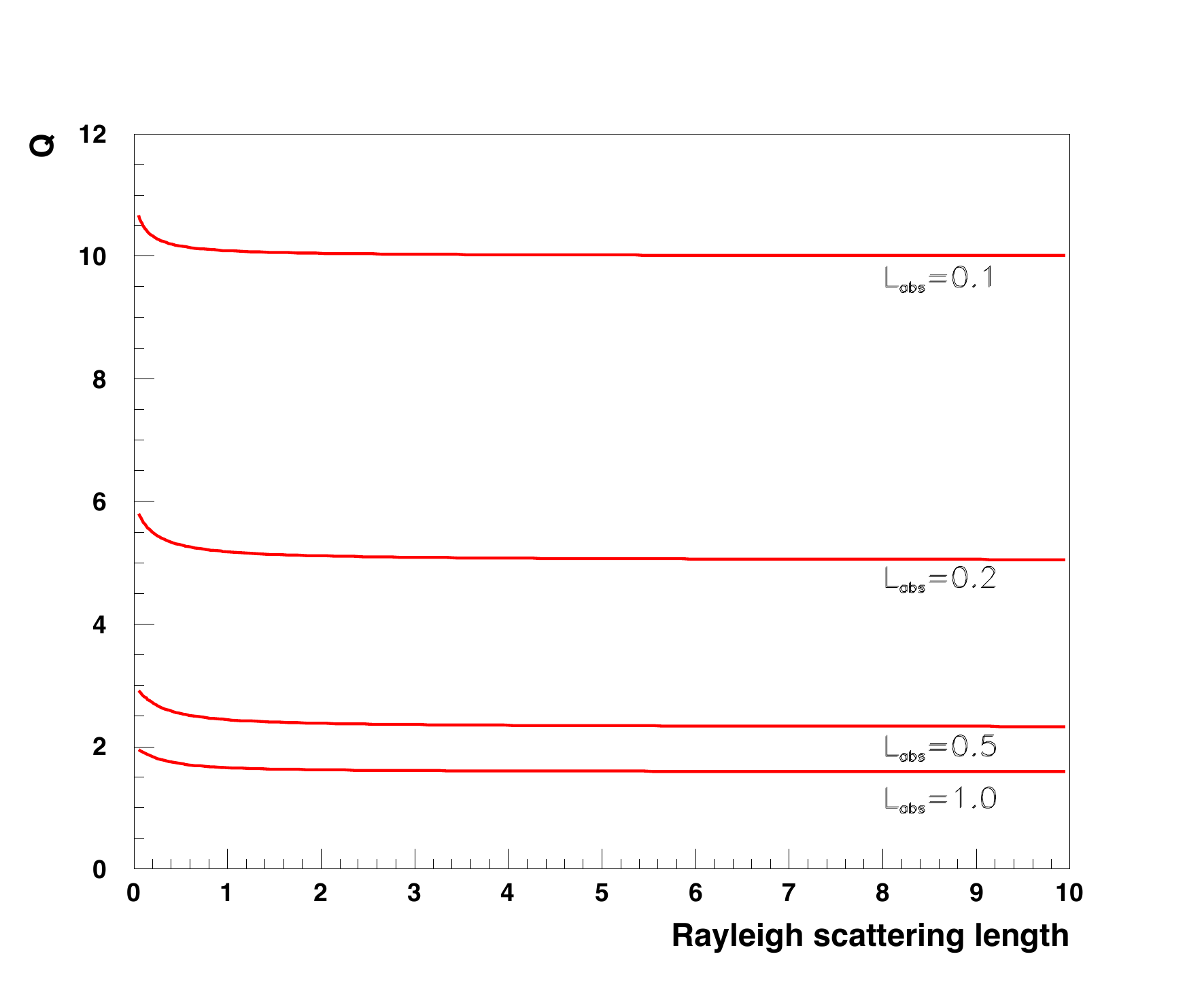}
\caption{\textsf{\textit{The term Q as a function of the Rayleigh scattering length in units of $\tilde{L}$ 
for few values of the absorption length $L_{abs}$ (also in units of $\tilde{L}$).}}}
\label{figisoab}
\end{center}
\end{figure*}
%**********************************************************************%

where $\tilde{L}$ is some characteristic linear dimension of the detector and $P(x)$ is the probability density distribution of the distances that a photon would travel in absence of 
interactions between two reflections (or between one reflection and absorption/detection). 
%We assume that in case one of the two points is chosen inside the active medium the same $P(x)$ 
 To be consistent with the case of 
regular solids one can define $\tilde{L}=6V/S$ with $V$ the
volume and $S$ the boundary surface of the detector\footnote{This gives $\tilde{L}=2R$ for the sphere, $\tilde{L}=L$ for a cube and $\tilde{L}=2R$ for a square cylinder.} and as a
first order approximation one can choose $P(x)$ to be uniform between $0$ and $\tilde{L}$, so that:

\begin{equation}
\label{eq:Ura_ex}
U_{RA}=\int_{0}^{\tilde{L}}\frac{1}{\tilde{L}}e^{-\frac{x}{\tilde{\lambda}}}dx = \frac{\tilde{\lambda}}{\tilde{L}}~(1-e^{-\frac{\tilde{L}}{\tilde{\lambda}}})
\end{equation}

and $Q$ is obtained by substituting this value to $U_{RA}$ in equation \ref{eq:Q}. In figure \ref{figisoab}  it is shown the plot of the term $Q$  as a function of the Rayleigh scattering
length for fixed values of the absorption length, both in units of $\tilde{L}$. $Q$  is only weakly dependent on the normalized Rayleigh  scattering  length and visible effects can be 
seen only when it is smaller than one.\footnote{It is implicit that the Rayleigh scattering can influence  absorption only if the absorption length is different from zero, otherwise it has 
no effect (Q=1).} On the other side the dependence on the (normalized) absorption length is much stronger and for $L_{abs}=1$ the term  $Q$ is already near to 2.

\section{Monte Carlo tests of the model.}
The predictions of this simple toy model  have been compared with the results of Monte Carlo simulations of specific scintillation detectors. In order to directly use the formulas 
found for the example developed in section \ref{sec:basic} a scintillator contained in a cubic box is firstly considered. The cubic shape has been chosen  because it has a good 
degree of
symmetry, but at the same time is sufficiently  far from a sphere, that represents the best approximation of the hypotheses that have been used and thus can be a good benchmark
for the toy model.\\
%*****************************FIGURE 2*****************************%
\begin{figure*}[tbp]
\includegraphics*[width=7.3cm,height=5.cm]{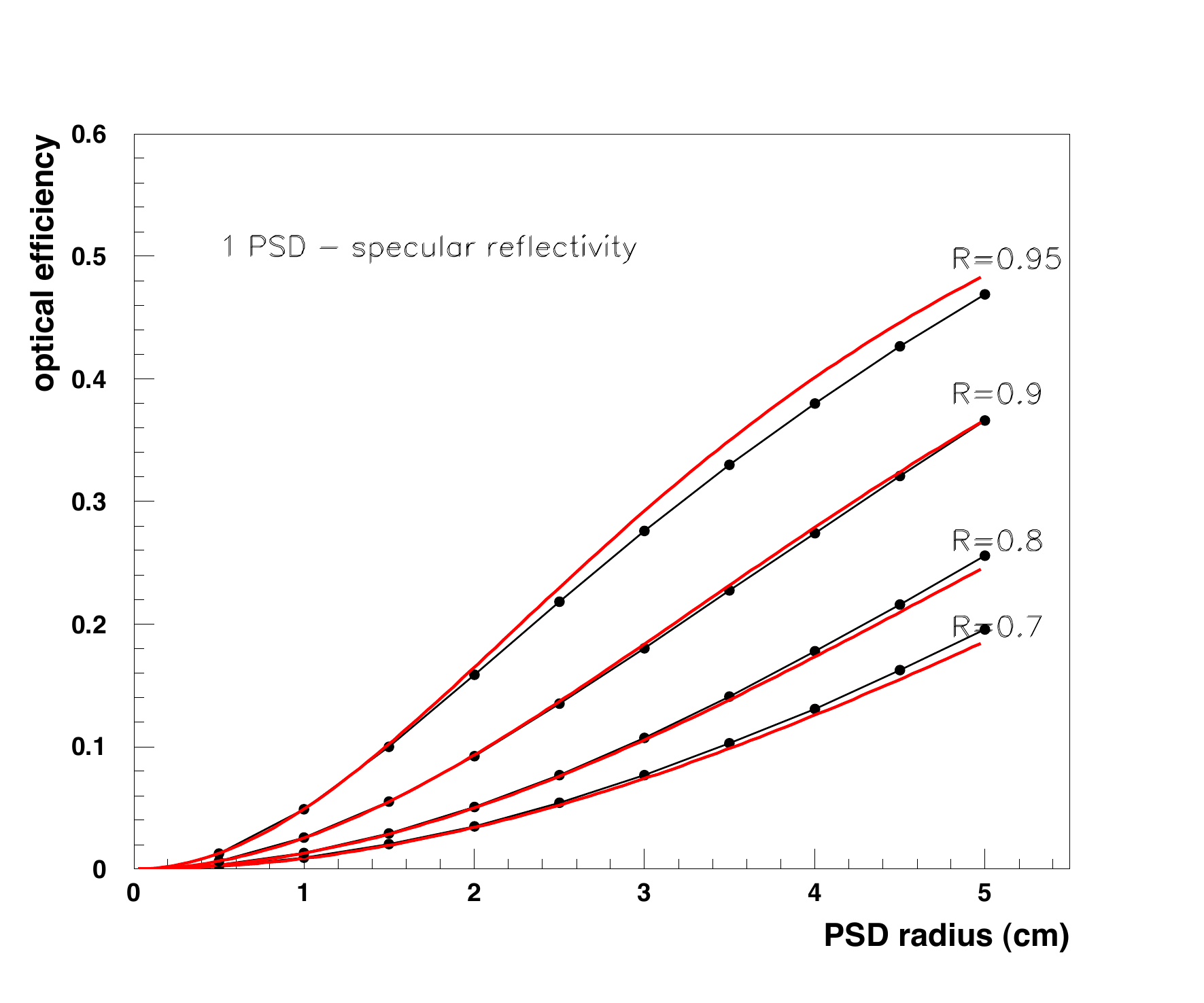}
\includegraphics*[width=7.3cm,height=5.cm]{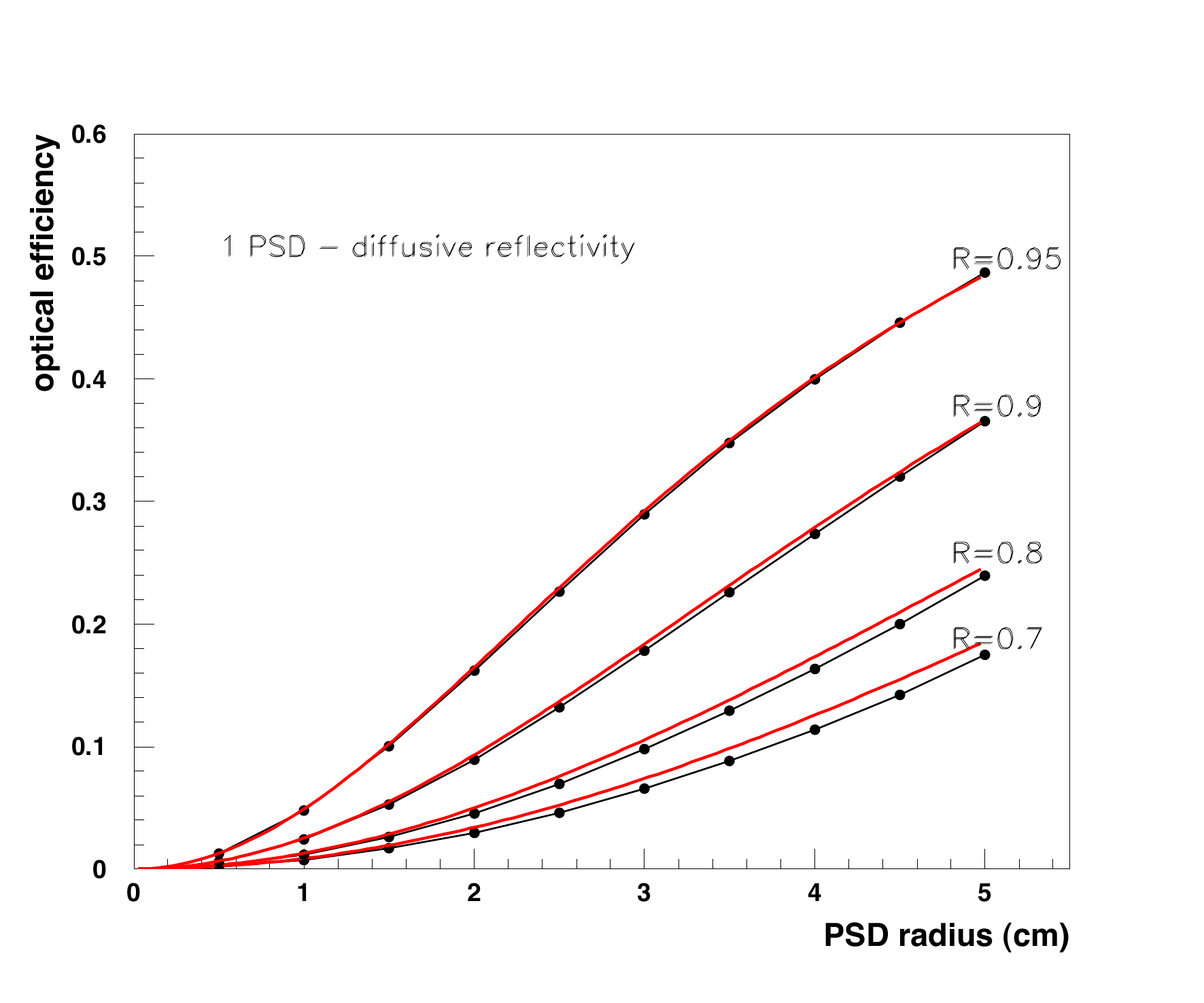}
\includegraphics*[width=7.3cm,height=5.cm]{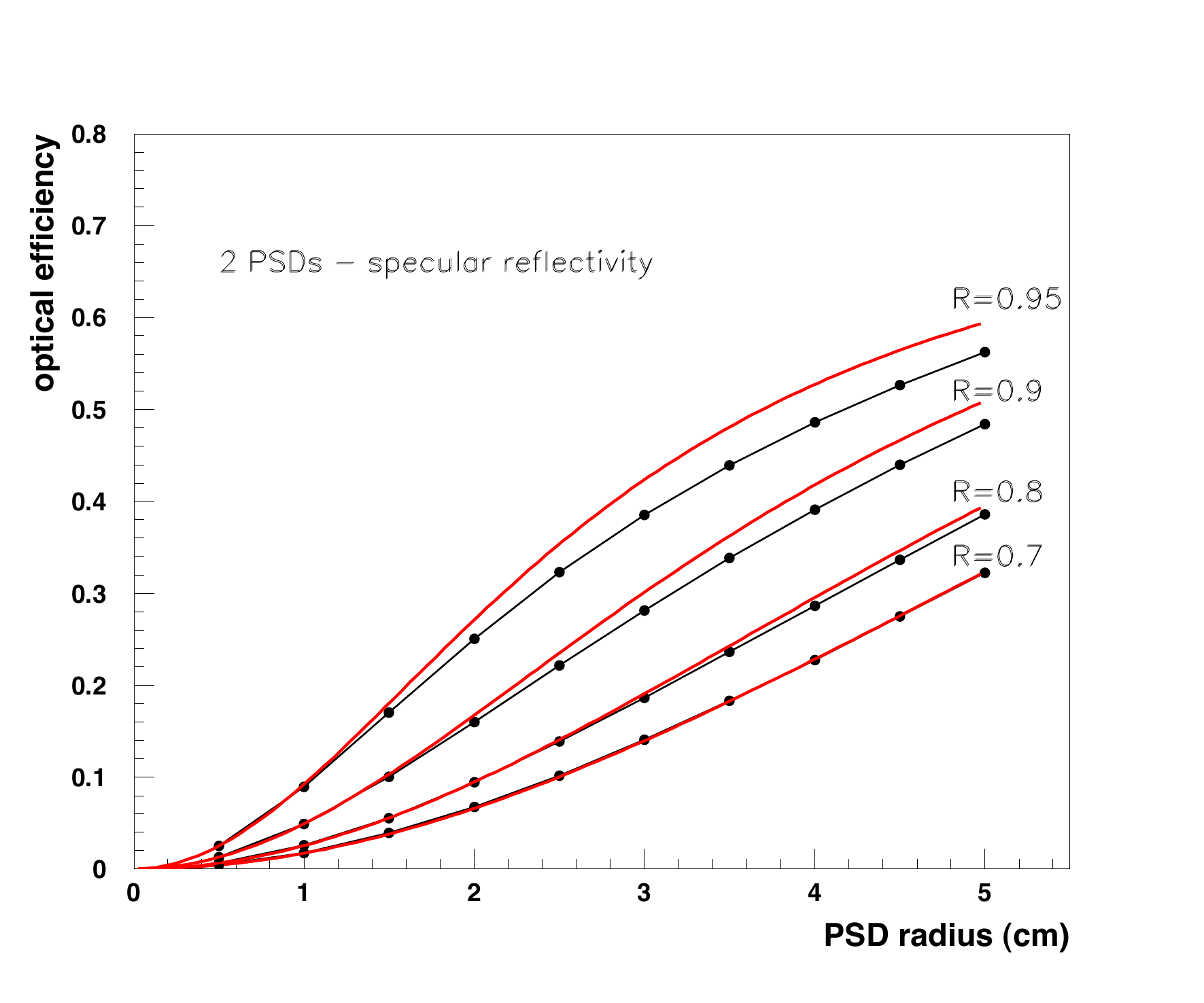}
\includegraphics*[width=7.3cm,height=5.cm]{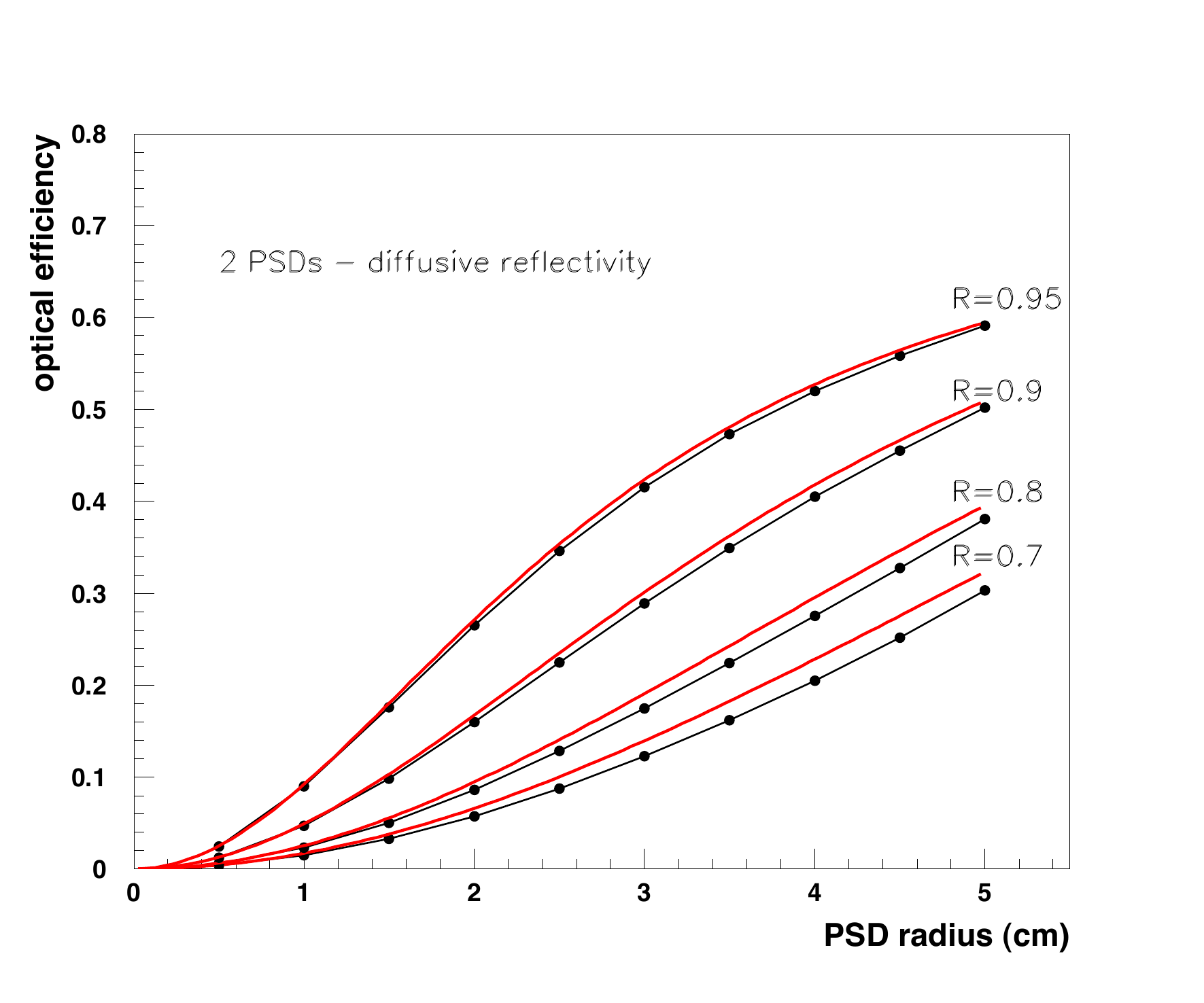}
\includegraphics*[width=7.3cm,height=5.cm]{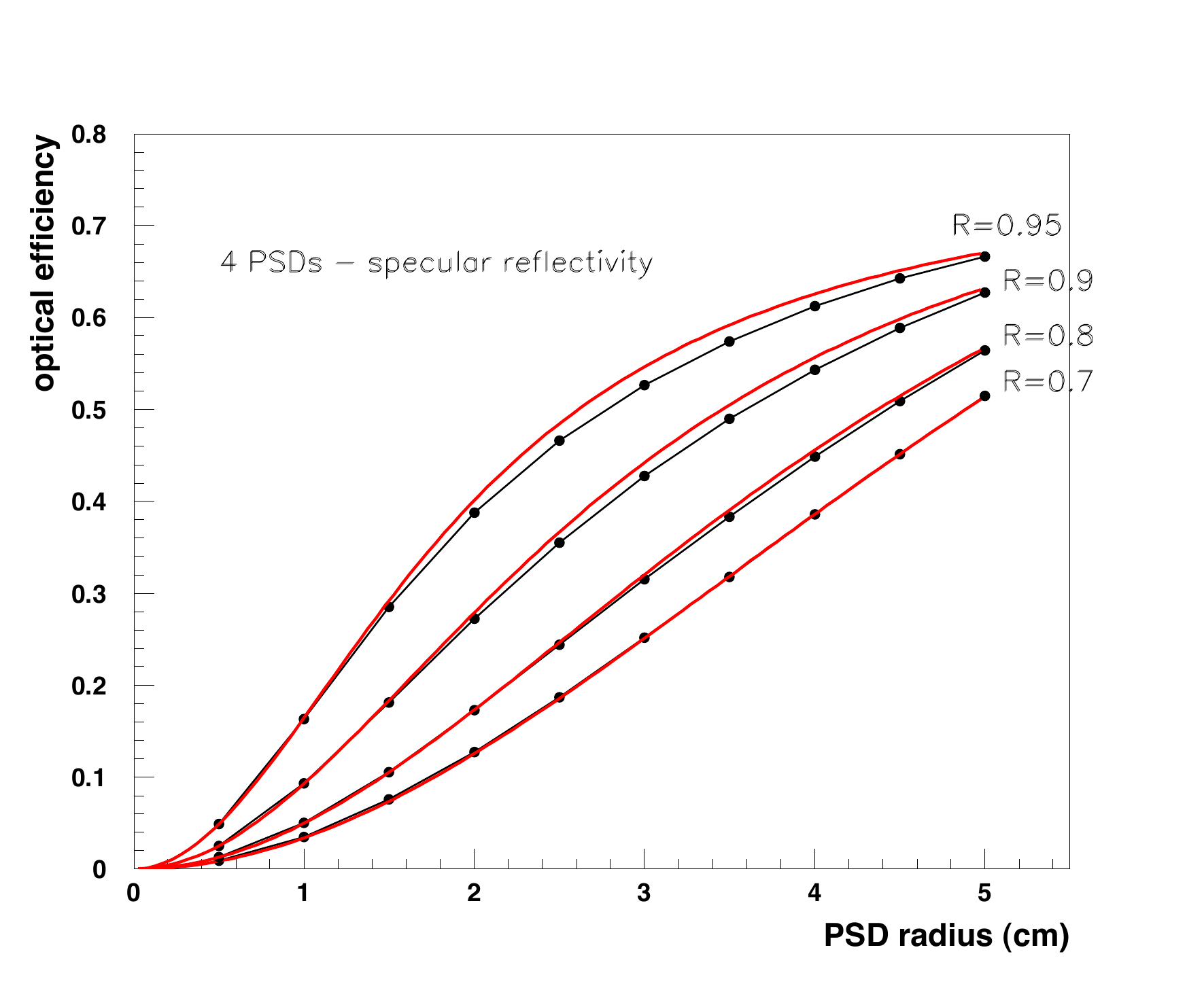}
\includegraphics*[width=7.3cm,height=5.cm]{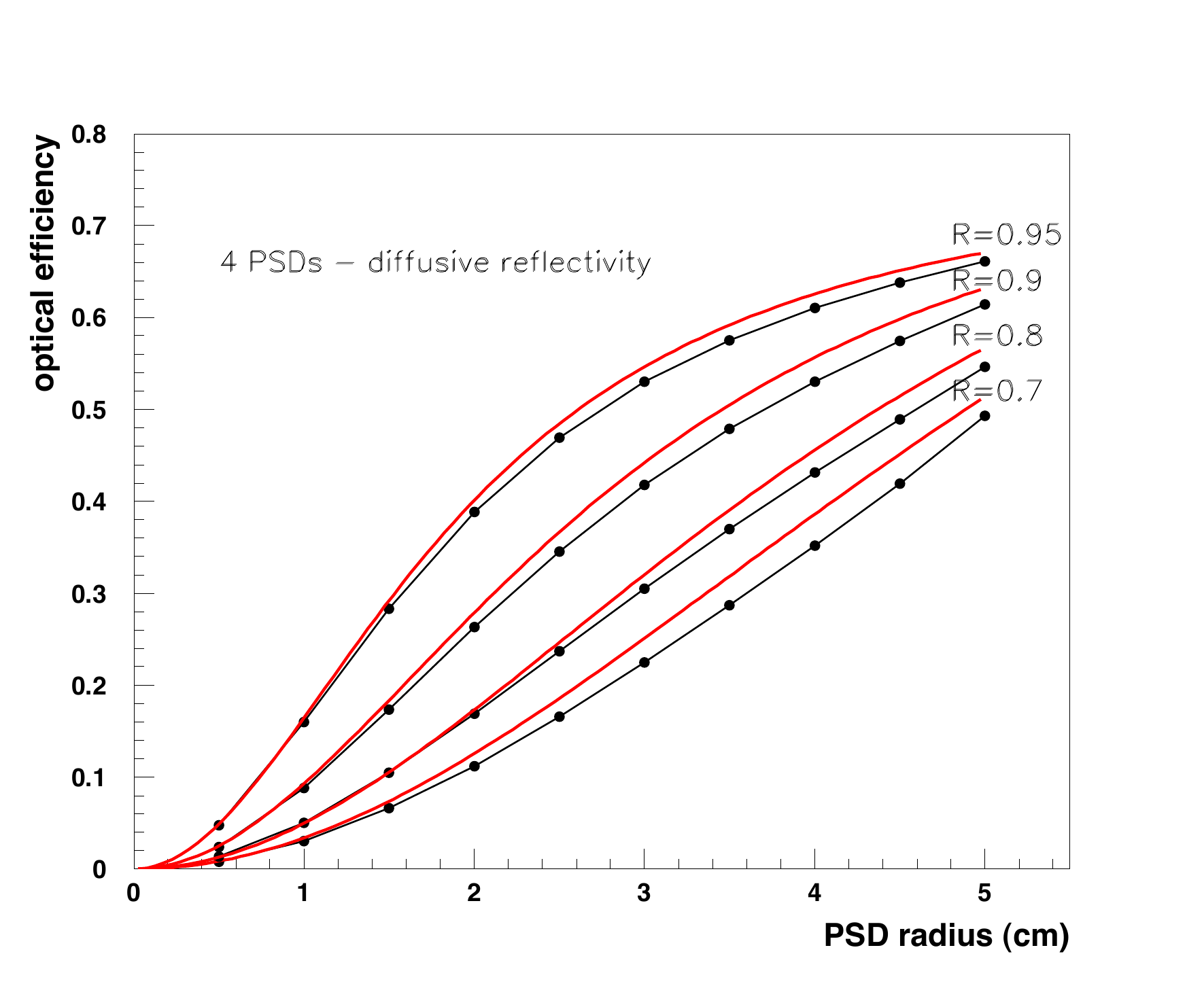}
\caption{\textsf{\textit{Top: Optical efficiency for the cubic scintillator with one (Top), two (Middle) and four (Bottom) PSD(s). Specular reflectivity on the left and diffusive reflectivity on
the right. Black dots represent the results of Monte Carlo simulations, while red lines are the model predictions.}}}
\label{fig1psd}
\end{figure*}
%**********************************************************************%

The cube is assumed to have a side of length $L=10~cm$ and is observed by one PSD with circular flat window. The PSD window is positioned exactly in the middle of one of the
cube's faces and is assumed to have a reflectivity $R_{w}=0.3$ and a transmissivity $T_{w}=0.5$, while the reflectivity $R$ of the  internal non active surface is varied between 0.70 
and 0.95. The radius  of the PSD window is varied between $0.5~cm$ and $5.0~cm$.  The optical efficiency of the detector, $\epsilon_{opt}$,  for any given configuration of the 
parameters is evaluated by randomly extracting a point inside the cube and generating from it a huge number of photons ($10^5$) with direction uniformly distributed in space. This
procedure is repeated for $10^5$ times and each time the fraction $N_{det}$ of photons transmitted across the PSD window with respect to generated ones is stored.\\
The  average fraction of \emph{detected} photons is determined  by fitting the distribution of $N_{det}$ with a  Gaussian function and taking its central value. \\
The outcomes of the simulations are displayed as black dots in figure \ref{fig1psd}, Top. Two cases have been separately considered: the case of completely specular reflections
(left) and the case of completely diffusive (lambertian) reflections (right). To estimate the detection efficiency equation \ref{eq:simple_scint_1} has been used (red lines in figure 
\ref{fig1psd}), where $f$ is the fraction of the cube's surface occupied by the PSD window.\\ 
The simulation has been repeated for the same cubic cell but with \emph{two} and  \emph{four identical PSDs}  installed on different non-adjacent faces of the cube. The results, 
together
with the predictions of equation \ref{eq:simple_scint_1} are shown in figure \ref{fig1psd} (Middle and Bottom respectively).\\
In the case of two PSDs,  equation \ref{eq:simple_scint_1} has  been computed for a single PSD with a reflectivity $\tilde{R}$ for the remaining internal surface of:

\begin{equation}
\label{eq:reflect_2}
\tilde{R} = \frac{(1-2f)R+ fR_{w}}{1-f}
\end{equation}  

The optical efficiency of the system is two times the one of the single PSD, since the two PSDs are identical.  In the case of four PSDs  an analogous calculation has been
performed. For all the examined cases small differences between specular and diffusive reflectivity are found. This simple model very well reproduces the results of the Monte
Carlo simulations and discrepancies at the level of few percent are found.\\

\begin{figure*}[tbp]
\begin{center}
\vspace{-2.0cm}
\includegraphics*[width=11.cm,height=9.cm]{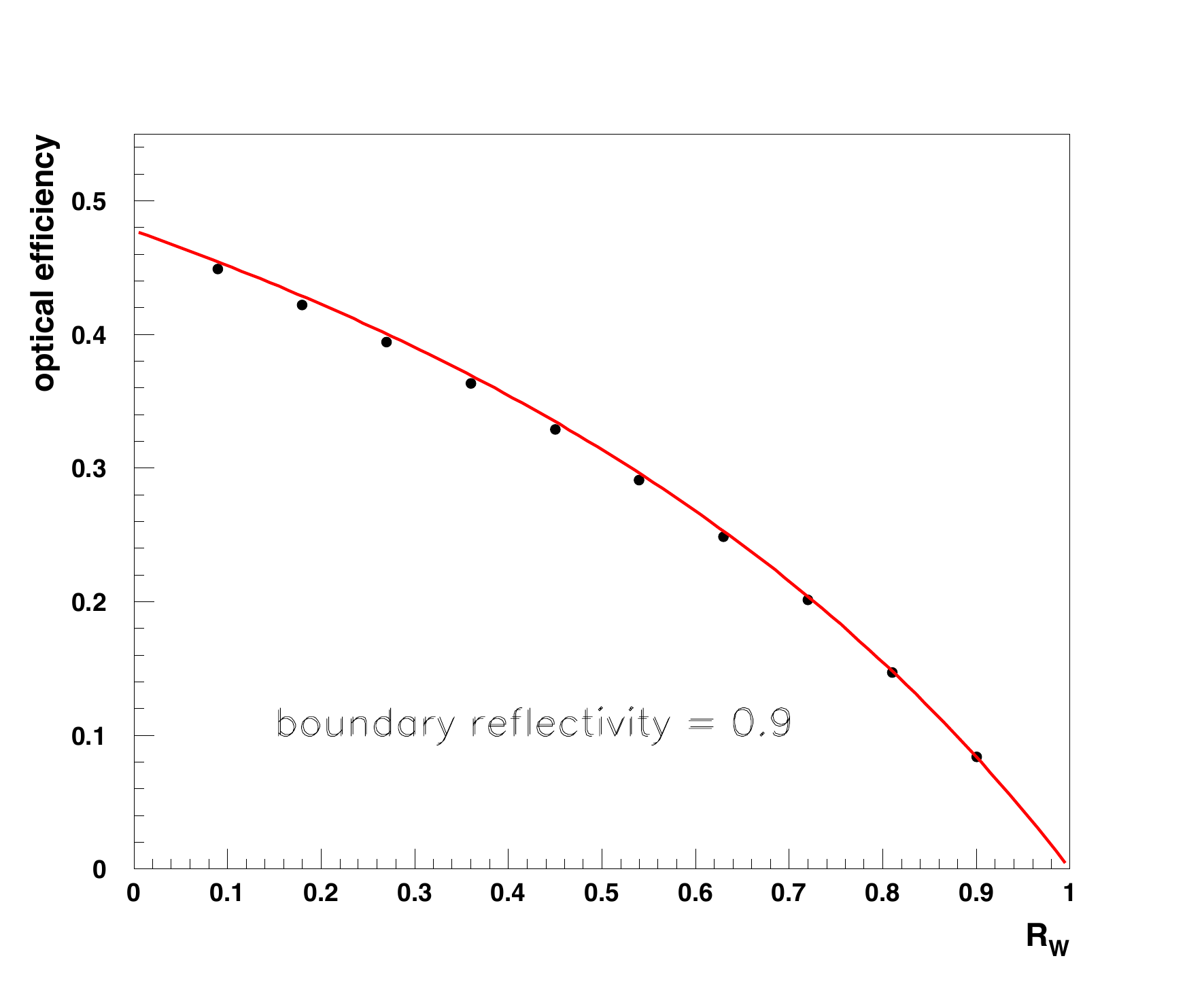}
\caption{\textsf{\textit{Dependence of the optical efficiency on the wall's reflectivity $R_{w}$. The cubic cell has one PSD and the reflectivity of the walls has been fixed at 90$\%$. 
The transmissivity of the window is $T_{w}=1-R_{w}$. Black dots comes from simulation and red line represents the prediction of the model.}}}
\end{center}
\label{figrwscan}
\end{figure*}

\begin{figure*}[tbpc]
\includegraphics*[width=7.3cm,height=5.cm]{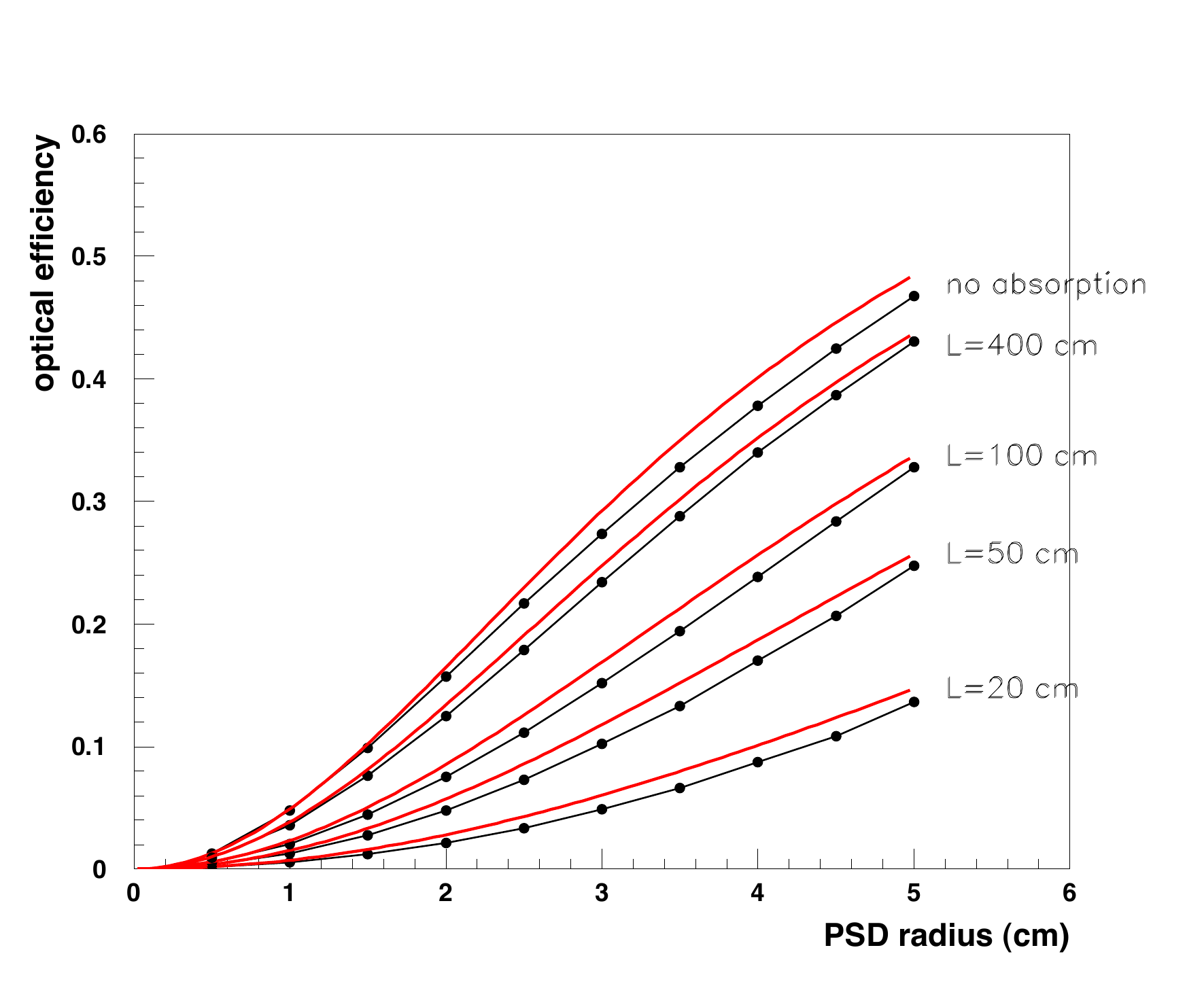}
\includegraphics*[width=7.3cm,height=5.cm]{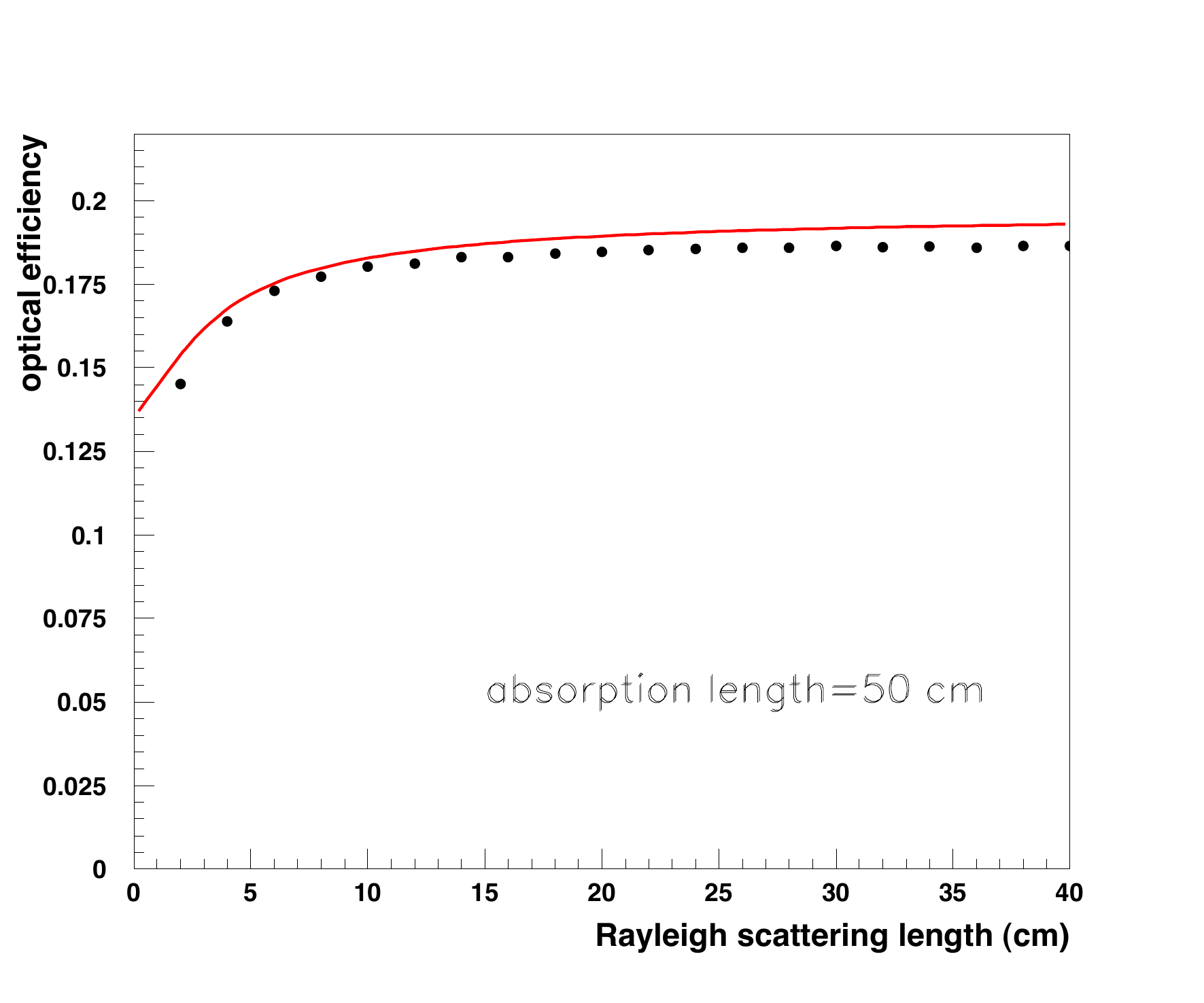}
\caption{\textsf{\textit{Left: optical efficiency of the cubic scintillator with one PSD. Rayleigh scattering length = 10 cm. Attenuation length varied between 10 and 400 cm. Right: 
optical efficiency as a function of the Rayleigh scattering length. The attenuation length is fixed at 50 cm, the PSD window has a radius of 4 cm and the reflectivity of the non 
instrumented surface is 0.95. Black dots represent the results of the simulation while the red line is the prediction of the model.}}}
\label{fig1psd_RA}
\end{figure*}

The dependence of the optical efficiency on the window's reflectivity $R_{w}$ has been tested with a dedicated  simulation of the cubic cell with one PSD. The PSD radius is fixed 
at $4~cm$, the (specular) reflectivity of the walls at 0.90 and $R_{w}$ is varied between 0.1 and 0.9. The transmissivity of the window $T_{w}$ is set at $1-R_{w}$. The results are
shown in figure \ref{figrwscan}. Also in this case equation \ref{eq:simple_scint_1} (red line) cleanly reproduces the outcomes of the simulation (black dots).\\
%*****************************FIGURE 3*****************************%
%**********************************************************************%
The predictions of equation \ref{eq:eopt_RA}, that is in the case Rayleigh scattering and absorption effects are present, have been tested again with  the cubic scintillator with one
PSD installed. In this case the (specular) reflectivity of  the non active surface is set at 0.95 and the circular window radius of the PSD  varied between 0.5 cm and 5 cm. 
As a first the Rayleigh scattering length (that typically has very small effects, see section \ref{sec:ray_abs}) is fixed at 10 cm and the absorption length is varied between 10 cm and 
400 cm. Results are shown in figure \ref{fig1psd_RA} left. With the same choice of parameters, except the absorption length and PSD radius fixed respectively at 50 cm and 4 cm, 
the dependence of the optical efficiency from the Rayleigh scattering length has been separately tested by varying it from 2 cm to 40 cm. Results are shown in figure 
\ref{fig1psd_RA} right. It is impressive how the term Q calculated from equations \ref{eq:Q} and \ref{eq:Ura_ex}, that are derived from very general considerations, accounts for the
effects of scattering and absorption over all the scanned lengths and with this level of accuracy.\\
  
The tests have been repeated with many different choices of the optical parameters (boundary surface reflectivity, PSD window dimensions, Rayleigh scattering length, absorption
 length, ...) and the agreement between MC data and model predictions has always been found at the level shown here.\\
%*****************************FIGURE 4*****************************%
%**********************************************************************%

\begin{figure*}[tbh]
\includegraphics*[width=7.3cm,height=5.cm]{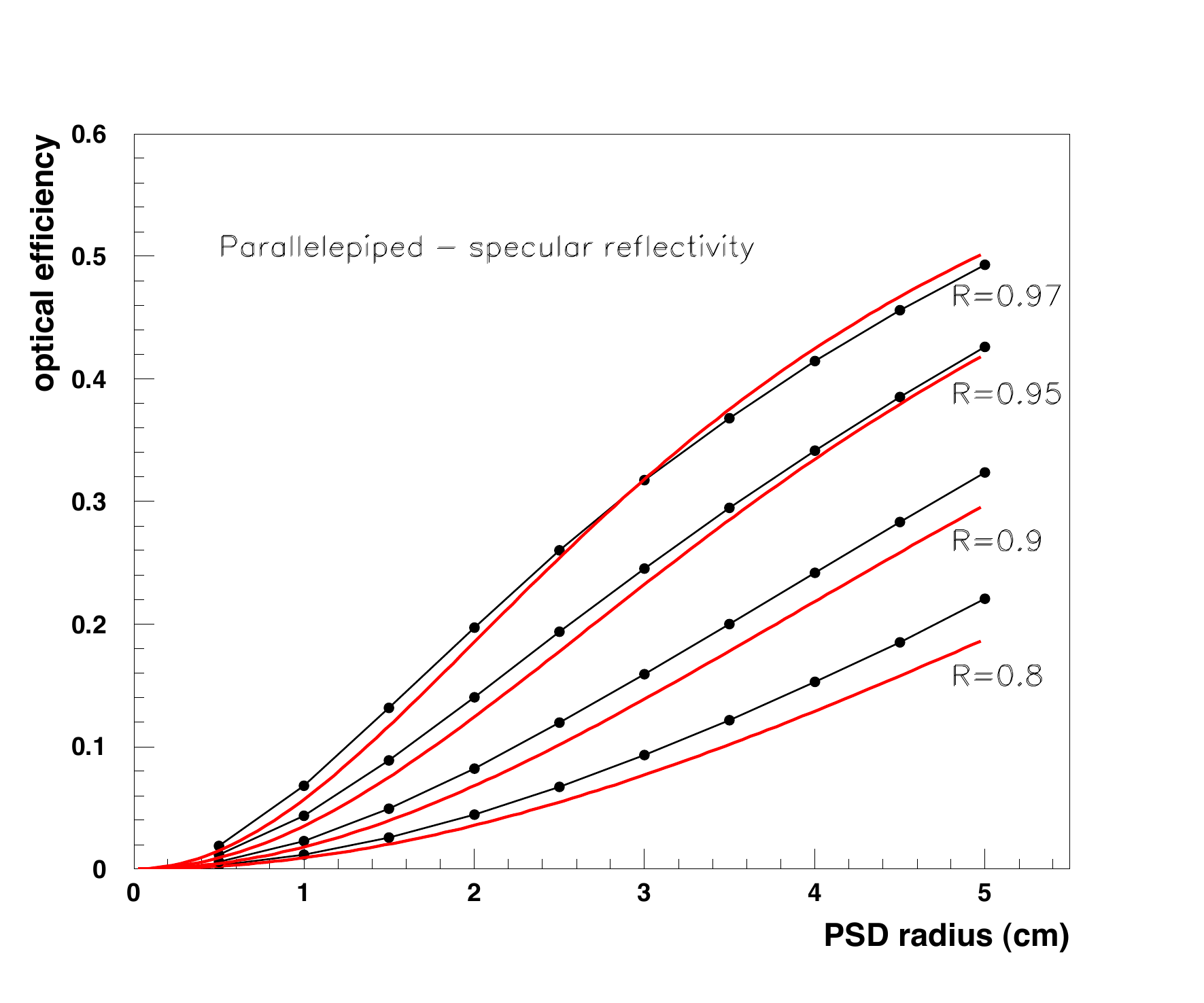}
\includegraphics*[width=7.3cm,height=5.cm]{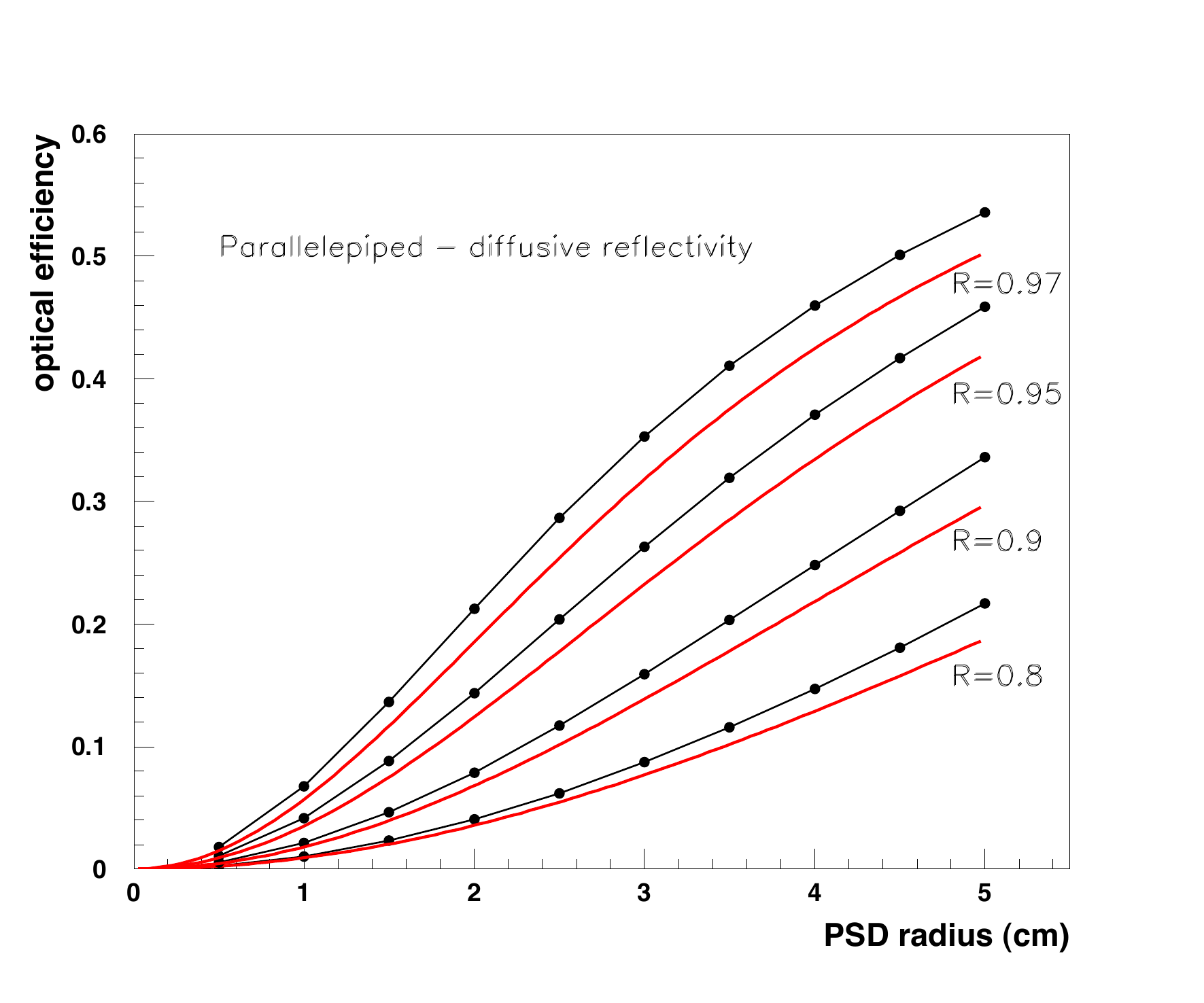}
\caption{\textsf{\textit{Optical efficiency of the parallelepiped scintillation detector.  Left: specular reflectivity. Right: diffusive lambertian reflectivity. Black dots represent the results of 
Monte Carlo simulations, while red lines are the model predictions.}}}
\label{figparalsim}
\end{figure*}

For the last and most challenging test a scintillation detector with a poor degree of symmetry is considered. The active medium is contained in a parallelepiped box with internal 
dimensions: 10 cm x 10 cm x 30 cm (l x w x h). Two identical PSDs with circular window are installed on the two opposite square (10 cm x 10 cm) faces of the box. The reflectivity of 
the passive internal surface is varied between 0.8 and 0.97 (specular or diffusive) and the radius of the PSD windows between 0.5 and 5 cm. The windows' reflectivity is set at 0.3 
and its transmissivity at 
0.5 (as for the cubic scintillator). Again the optical efficiency is evaluated with dedicated MC simulations and compared with the prediction of equation \ref{eq:simple_scint_1} ($f$ is 
the PSD coverage of the internal surface). The two cases of specular and diffusive reflectivity are treated separately. The results of these tests are shown in figure \ref{figparalsim}. 
Surprisingly also in this case the agreement between MC outcomes and model predictions is 
 satisfactory good. In fact, for specular reflectivity  above 0.9 discrepancies at the level of few percent are found, while in all other cases they are of the order of 10\%.
 
%*****************************FIGURE 5*****************************%
%**********************************************************************%   

\section{Adding a wavelength shifter to the scintillator.}
Some materials emit scintillation photons in the ultraviolet region of the electromagnetic spectrum. In these cases it is a common practice to dilute wavelength shifting substances 
(WLS) in the scintillator (\cite{borex}, \cite{lvd}) that absorb ultraviolet light and re-emit visible photons easily detectable with glass windowed photomultipliers. The formulas 
obtained in the previous sections for the evaluation of $\epsilon_{opt}$ are still perfectly applicable with the small modification of introducing  an overall efficiency for the conversion 
of photons (typically near to one). If it is not possible to dilute any WLS in the scintillator an alternative solution consists in depositing it (by vacuum evaporation, sprying, painting, ...)
on the internal surface of the detector, PSDs' windows included. This is the case, for instance,  of double phase argon detectors (\cite{warp_100}, \cite{ardm_prop}), where the 
stringent requests on the liquid purity limits the admissible amount of diluted contaminants to tens of ppb of electronegative substances and hundreds of ppb of non-electronegative 
ones (\cite{warp_N2}, \cite{warp_O2}). This situation can be well handled with the ideas developed above, but some particular care must be taken.\\
 
Consider a scintillation detector with only one PSD (situation easily generalizable to the case of  $n$ PSDs, as shown above). Assume that the scintillator medium is contained in a 
box of regular shape, so that equation \ref{eq:simple_scint_1}, generalized to the case of $Q\ne 1$, can be used. Define $\epsilon_{WS}$ as the shifting efficiency of the non 
instrumented  internal surface and assume that $R$ is their reflectivity to shifted photons. Define also $\epsilon_{ws}$ as the shifting efficiency of the  PSD window that has a
 transmissivity $T_{w}$, a reflectivity $R_{w}$ (to shifted photons) and that covers a fraction $f$ of the internal surface of the cell.\\

The crucial point is here the  calculation of the probability that a VUV (Vacuum Ultra Violet) photon reaches the boundary surface of the detector where it is wavelength-shifted. A
simple argument (see Appendix A) shows that this probability is equal to the inverse of the term Q  (section \ref{sec:ray_abs}) calculated using the Rayleigh scattering length and 
the absorption length of  VUV photons in the scintillator medium ($Q_{VUV}$). To evaluate $\epsilon_{opt}$ it is necessary to consider that:  

\begin{itemize}
\item the probability that a  VUV photon reaches the boundary surface of the cell is $\frac{1}{Q_{VUV}}$;
\item the probability that the photon is  down-converted on the window of the PSD is $\frac{f\epsilon_{ws}}{Q_{VUV}}$. Since the emission process is isotropic, the probability that it 
is \emph{directly} transmitted across the PSD window is roughly half of the total: 
\begin{equation}
\epsilon_{opt}^{direct}=\frac{f\epsilon_{ws}}{2Q_{VUV}} 
\end{equation}
eventually reduced by a factor that takes into account the absorption of the window. The complementary (half) part is the probability that the photon is sent back in the cell;
\item the probability that the photon is  down-converted on the non instrumented surface of the cell is $\frac{(1-f)\epsilon_{WS}}{Q_{VUV}}$. The shifted photon will propagate inside
the cell and will be detected with a probability $F(Q_{vis},T_{w}f,R(1-f)+R_{w}f)$. Where $Q_{vis}$ is calculated using Rayleigh scattering and absorption lengths for \emph{visible} 
photons; 
\item the probability that the photon is  \emph{indirectly}  detected is then: 
\begin{equation}
\epsilon_{opt}^{indirect}=\Big(\frac{f\epsilon_{ws}}{2Q_{VUV}}+\frac{(1-f)\epsilon_{WS}}{Q_{VUV}}\Big) F(Q_{vis},T_{w}f,R(1-f)+R_{w}f) 
\end{equation}
Here we take into account that the photon can come form the PSD's window or form the inactive surface.
\end{itemize}
 
In conclusion the total detection probability is:
\begin{equation}
\epsilon_{opt}=\epsilon_{opt}^{direct}+\epsilon_{opt}^{indirect}=\frac{T_{w}f\epsilon_{ws}}{2Q_{VUV}}+\Big( \frac{f\epsilon_{ws}}{2Q_{VUV}}+\frac{(1-f)
\epsilon_{WS}}{Q_{VUV}}\Big) F(Q_{vis},T_{w}f,R(1-f)+R_{w}f)
\label{eq:epsilon_shifted}
\end{equation}
It is interesting to notice here that because of the term $\frac{(1-f)\epsilon_{WS}}{Q_{VUV}}$ this is not a priori a monotonically increasing function of the PSD coverage $f$. It is then 
possible that for a certain set of optical parameters the maximum value of the optical efficiency is not reached with a total PSD coverage of the internal surface, i.e. $f=1$, but for 
some optimal value that can be found by maximizing $\epsilon_{opt}$ (equation \ref{eq:epsilon_shifted}) with respect to $f$.\\
 
  %*****************************TABLE 2*****************************%
\begin{table}[htbp]
\begin{center}
\caption{\textsf{\textit{Parameters used to evaluate the LY of the detector described in \cite{ly_warp_paper}.}}}
\vspace{0.5cm}
\begin{tabular}{l l} 
\hline
 \hline  
photon yield & $N_{\gamma}$=40 photons/keV \cite{doke}\\
photocathodic coverage & f=13\%\\
transmissivity of PMT window & $T_w$=0.94 \cite{ham_hand}\\
reflectivity of PMT window & $R_w$=0\\ 
conversion efficiency of PMT & $\epsilon_{PSD}$= 28\%\\
no absorption of VUV photons & $Q_{VUV}$=1\\
no absorption of visible photons & $Q_{vis}$=1\\
conversion efficiency of passive surface & $\epsilon_{WLS}$=1 \cite{TPB_prop}\\
conversion efficiency of PMT window\\(no shifter)&$\epsilon_{wls}$=0.\\
reflectivity of passive surface (reflector+TPB) & R=0.95 \cite{ENEA}\\
\hline
\hline
\label{tab:0.7l}
\end{tabular}
\end{center}
\end{table}
%**********************************************************************%  
 
Equation \ref{eq:epsilon_shifted} has been used to evaluate the LY of the scintillation detector described in full detail in \cite{ly_warp_paper}. The scintillating medium is 
liquid  Argon that is contained in a cylindrical PTFE cell (h=9.0 cm and $\phi$=8.4 cm) and is observed by a single 3" photomultiplier. Liquid Argon is an abundant scintillator 
($\sim$ 40 photons/keV) but photons are emitted in the VUV region of the electromagnetic spectrum ($\lambda$=128 nm) and need to be wavelength shifted to be detected with 
the installed photomultiplier (synthetic silica window - cutoff around 200 nm). For this reason the internal surface of the cell is completely covered with a reflective foil deposited with 
Tetra Phenyl Butadiene (TPB),  that is an extremely efficient shifter with emission spectrum peaked around 420 nm  \cite{TPB_prop}  \cite{ENEA}. The parameters used to evaluate 
the LY of the detector are summarized in table \ref{tab:0.7l}.

For $\epsilon_{PSD}$  the product of the photocathode quantum efficiency averaged over the TPB emission spectrum (29.5\%) and of the photoelectron's  collection efficiency at 
first dynode (95\%) has been taken.  Given the refractive index of LAr  ($n_{LAr}$(420 nm)=1.25 \cite{cerenkov}) and of the synthetic silica window ($n_{window}$(420 nm)=1.46
 \cite{fused_silica}) reflection of photons at the interface can be neglected in good approximation, and hence $R_w$ has been set to zero.
%$R_w$ has been set to zero because the refractive index of liquid Argon for shifted photons ($n_{LAr}$(420 nm)=1.25 \cite{cerenkov}) is lower 
%than the refractive index of the synthetic silica window ($n_{window}$(420 nm)=1.46 \cite{fused_silica}) and no total photons' reflection can happen on the interface.\\ 
The LY resulting from equations \ref{eq:ly_bare} and \ref{eq:epsilon_shifted} is 6.9 phel/keV, in perfect agreement with the measured value of 7.0 phel/keV $\pm$ 5\%.

\section{Conclusions.}
A toy-model for the estimation of the light yield of a scintillation detector based on very simple hypotheses has been developed. It has been shown how to 
include  the effects related to Rayleigh scattering and absorption of the photons.\\
The model has been benchmarked with the outcomes of the Monte Carlo simulation of a cubic scintillator observed by one, two or four PSDs and has  demonstrated an accuracy at
the level of few percent.
An additional Monte Carlo test with a parallelepiped detector with a poor degree of symmetry has been performed. Even in this case the model predictability has resulted 
surprisingly good and an agreement with Monte Carlo outcomes better than 10\% has been obtained.\\
The model has been also applied to the estimation of the light yield of a real liquid Argon scintillation detector and  a value of 6.9 phel/keV has been 
found, perfectly compatible with the measured value of 7.0 phel/keV $\pm$ 5\%.\\ The formulas here reported can be adequate in all those cases a quite robust estimation of the 
light yield of simple scintillation detectors is needed. It can result very useful in the optimization of the  design of the detector since the dependence of the light yield from the optical 
parameters is completely explicit. Even in presence of a Monte Carlo simulation of the detector the model can be useful to cross-check and validate its predictions.\\
Furthermore the formulas found for the examples treated along the paper can be directly used in many real applications.

\appendix
\section{VUV photons absorption.}
Consider a scintillation detector and assume that photons suffer Rayleigh scattering and absorption. The optical efficiency of the detector is (equation \ref{eq:eopt_RA}):
\begin{equation}
\epsilon_{opt}=\frac{\alpha_{0}}{Q-\beta_{0}} 
\end{equation}
where $\alpha_0$ and $\beta_0$ are the detection and regeneration probabilities in absence of scattering/absorption. If, unlike what has been done in section \ref{sec:ray_abs}, 
the step definition is not changed and remains as the photon's propagation between two subsequent reflections,  the detection probability will be $q\alpha_0$ and the regeneration 
probability $q\beta_0$, where $q$ is the photon's surviving probability along the step. Consequently the optical efficiency can be written as:
\begin{equation}
\label{eq:phabs}
\epsilon_{opt}=\frac{q\alpha_0}{1-q\beta_0}=\frac{\alpha_0}{1/q-\beta_0}
\end{equation} 

Comparing equations \ref{eq:eopt_RA} and \ref{eq:phabs}:
\begin{equation}
q=\frac{1}{Q}
\end{equation}\\

\end{document}